\documentclass[11pt,notoc]{JHEP3}

\usepackage{epsfig}
\usepackage{mathrsfs}
\usepackage{amsmath,amssymb}
\usepackage{graphicx}

\newcommand{\bea}{\begin{eqnarray}}
\newcommand{\eea}{\end{eqnarray}}
\newcommand{\p}{\partial}
\newcommand{\LL}{\mathcal {L}}
\newcommand{\ov}{\over}
\newcommand{\dd}{\textrm{d}}

\title{Holographic melting of Heavy Baryons in Plasma with Gluon Condensation}

\author{ Sang-Jin Sin\footnote{sangjin.sin@gmail.com}\\
Department of Physics, Hanyang University, Seoul 133-791, Korea\\
}

\author{ Yang Zhou\footnote{yzhou@itp.ac.cn}\\
Institute of Theoretical Physics\\
Chinese Academic of Science\\
Beijing 100190, PRC\\

}
 \abstract{We propose a Dirac-Born-Infeld ( DBI ) D5 vertex brane plus $N_c$ fundamental strings configuration to describe a
 baryon probe in strongly coupled gauge theory with gluon condensation at finite temperature via AdS/CFT correspondence.
 We investigate properties of this configuration in a dilaton deformed AdS$_5\times$S$^5$ background, in which IIB string
 theroy is dual to super Yang-Mills thery in a state with a constant self-dual gauge field ($F_{mn}=F^*_{mn}$) background.
 We find that for most values of temperature $T$ and gluon condensation parameter $q$ ($q = \pi^2<F_{mn}F_{mn}>$), there always
 exists a screening length $L_s$. The relation $L_s\sim{1\ov T}$ has been checked. We give the $q$ dependence of $L_s$.
 We calculate the boost velocity $v(v=-\tanh\eta)$ and angular velocity $\omega$ dependence of
$L_s$ for a baryon probe, and obtain $L_s=L_0*(1-v^2)^{1/4}$ for large $v$ and $L_s\sim\omega^{-1}$, which are
consistent with those dependence relations in the point brane plus strings case, and find that the usual relations have
been modified by $q$. We also calculate the mass of baryon and find $T$ dependence of baryon mass.}


\begin{document}

\tableofcontents

\section{Introduction and Summary}
There has been much interest in studying hadrons in strongly coupled QCD in terms of AdS/CFT correspondence. One
interesting topic is the investigation of holographic baryons~\cite{GI0801,SS0802,Ghoroku:2008na,Sfetsos:2008yr}. A
holographic baryon model in gauge/gravity duality was first introduced by Witten, where a baryon is identified with a
compact D brane wrapped on a transverse sphere with $N_c$ fundamental strings attached to it. Investigations of baryons
in AdS/CFT has been started ten years ago, in order to find a solution with a compact vertex brane and
Dirac-Born-Infeld strings, but there are many challenges~\cite{JLR:0606,HP:0405,Ima:0410,CGST:9902,CGST:9810}. One big
problem is how to get a closed brane solution for a baryon vertex from the DBI + CS(Chern-Simons) action in a certain
gravity background. We construct a new configuration with a wrapped D5 vertex brane and $N_c$ strings under the
background with gluon condensation, to solve these problems. We investigate properties of this configuration in a
dilaton deformed AdS$_5\times$S$^5$ background, in which IIB string
 theroy is dual to super Yang-Mills thery in a state with a constant self-dual gauge field ($F_{mn}=F^*_{mn}$) background.
The gluon condensation parameter $q$ appears in our dilaton and can be related to the self-dual gauge field by $q =
\pi^2<F_{mn}F_{mn}>$. In the string side, this dilaton comes from D-instantons homogeneously distributed over D3-brane
world-volume. More descriptions of this solution can be found in~\cite{Liu:1999fc}. In particular, we investigate
properties of this configuration in a probe limit and argue that this configuration may describe a heavy baryon well in
a certain plasma background.

Recently, authors of the work\cite{Liu:2008bs} proposed a simple configuration of baryon in a hot strongly coupled
super Yang-Mills plasma, analyzed the velocity dependence of baryon screening length $L_s$ and found that there was a
relation $L_s\sim L_0*(1-v^2)^{1/4}$ for baryon when $v$ went to one, which has been found in quark and anti-quark
screening~\cite{Liu:2006sw}. Furthermore, screening length and $J-E^2$ behavior ($J$ is angular momentum and $E$ is
baryon mass) of high spin baryons were analyzed~\cite{Li:2008py}. All these investigations are in the framework of
thermal super Yang-Mills gauge theory/AdS balck hole duality and component quarks are considered as probes.  Many
results of these investigations are similar to those of meson case~\cite{Liu:2006sw}, because there the vertex brane is
treated as a massive point in AdS$_5$, with an action depending only on the gravity potential.

In general, the vertex brane is not a point in AdS space, but an extended object. It is difficult to find a closed
vertex brane in many gravity backgrounds, especially for D5 vertex brane. Baryon configurations with D4 vertex brane
and $N_c$ fundamental strings were investigated very recently in work~\cite{Zhou:2008vf} and there were many
interesting properties. Recently a closed baryon vertex was found in the D3 branes background with gluon
condensation~\cite{Ghoroku:2008na}. We propose a full baryon probe configuration in this background in the present
paper and find many interesting properties of this configuration.

The present paper is organized as follows. In Sec.2, we construct the general baryon configurations. We analyze the
different baryon vertex solutions from the DBI + CS action of D5 compact brane and give the force balance condition
between the D5 vertex and $N_c$ fundamental strings. In Sec.3, we give a general analysis of screening length in a
standard way. We find that for most values of temperature $T$ and gluon condensation parameter $q$, there always
 exists a screening length $L_s$. The relation $L_s\sim{1\ov T}$ has been checked. We also give the $q$ dependence of $L_s$.
 We also calculate the boost velocity $v=-\tanh\eta$ and angular velocity $\omega$ dependence of
$L_s$ for a baryon probe, which are consistent to those dependence relations in the point brane plus strings case, and
find that the usual relations have been modified by $q$. We also calculate the mass of baryon and find $T$ dependence
of baryon mass. We give our conclusion and discussion in Sec.4.
\section{General baryon configurations}
The baryon construction in gravity involves $N_c$ fundamental strings with the same orientation, beginning at the heavy
quarks on the flavor brane and ending on the baryon vertex in the interior of bulk geometry, which is a D5 brane
wrapped on the S$^5$ ( AdS$_5\times$S$^5$ background ). Generally, $N_c$ quarks are allowed to be placed at arbitrary
positions in $\vec x$ space on the boundary. Note that these quarks are heavier than component quarks of mesons, and
the $N_c$ quarks bound states can not easily be considered as an effective field on the boundary ( fluctuations of
flavor brane in picture with flavor ).

The gravity theory dual to the thermal four-dimensional gauge theory is a solution of 10D Type-IIB supergravity under
the Freund-Rubin ansatz for self-dual five form field strength~\cite{Kehagias:1999iy,Liu:1999fc,Ghoroku:2005tf}. In
string frame, the solution can be written as follows
\bea
\begin{split}
 e^{-{1\over 2}\phi}\textrm{d}s_{10}^2&=
 -{r^2\over r_+^2}\biggr(1-{r_0^4\over r^4}\biggr)\dd t^2+{r^2\over r_+^2}\dd x_i\dd x^i+{1\over 1-{r_0^4\over r^4}}{r_+^2\over r^2}\dd r^2+r_+^2\dd\Omega_5^2,\\
\end{split}\eea with a dilaton  and an axion \bea e^\phi=1+{q\over r_0^4}\log{1\over 1-{r_0^4\over r^4}},\quad
\chi=-e^{-\phi}+\chi_0,
\eea
 where $i=1,2,3$ and $q$ is gauge fields condensate parameter~\cite{Liu:1999fc,Ghoroku:2005tf}. $\phi$ and $\chi$ denote the dilaton and
the axion respectively, and no other field configurations are considered here. This metric includes an AdS black hole
times a five-dimensional sphere, and the dilaton and axion depending on $r$. $r_+$ is the curvature radius of the AdS
metric, $r$ is the coordinate of the fifth dimension of AdS$_5$ and $r_0$ is the position of black hole horizon. The
temperature of the gauge theory is given by Hawking temperature of the black hole, $T={r_0\over \pi R^2}$. By duality,
the gauge theory parameters $N_c$ ( color number ) and $\lambda$ ( t'Hooft coupling constant ) are given by \bea
 \sqrt{\lambda}={r_+^2\over \alpha'}, \quad\quad {\lambda\over N_c}=g_{YM}^2=4\pi g_s,
\eea where ${1\over 2\pi\alpha'}$ is string tension and $g_s$ is the string coupling constant. The self-dual
Ramond-Ramond field strength is \bea
 F_{(5)}=\dd C_{(4)}=4r_+^4\Omega_5\dd\theta_1\wedge...\wedge \dd\theta_5-4{r^3\over r_+^4}\dd t\wedge \dd x_1\wedge \dd x_2\wedge \dd x_3\wedge dr,
\eea where $\Omega_5=\sin^4\theta_1\sin^3\theta_2\sin^2\theta_3\sin\theta_4$. The D5 brane carries a radial U(1) flux
and wraps the S$^5$ with radial extension. The action of D5 brane includes DBI action plus Chern-Simons action, given
by \bea
\begin{split}
 S_{D5}&=-T_5\int \dd^6\sigma e^{-\phi}\sqrt{-\det(g_{ab}+2\pi \alpha'F_{ab})}+T_52\pi\alpha'\int A_{(1)}\wedge
 \mathscr{P}(F_{(5)}),\\
  \end{split}
\eea where the 6D world volume induced metric $g_{ab}=\p_aX^\mu\p_bX^\nu G_{\mu\nu}$, and the pull back of five form
$\mathscr{P}(F_{(5)})=\p_{a_1}X^{\mu_1}...\p_{a_4}X^{\mu_5}F_{\mu_1...\mu_5}$. The D5 brane tension
 $T_5={1\over g_s(2\pi)^5l_s^6}$, and the world volume field strength of U(1) flux $F_{(2)}=\dd A_{(1)}$.
The Chern-Simons term endows D5 brane with U(1) charge. By the following consistent ansatz that describes the embedding
D5 brane \bea
 \tau=t,\quad\sigma_1=\theta, \quad\sigma_2=\theta_2,...\sigma_5=\theta_5,\quad r=r(\theta),\quad
 x=x(\theta),
 \eea 
 we see only SO(5) symmetric configurations of D5 brane which stand for baryons in 4D real spacetime ($t,\vec x$) are considered,
 and the embedding function can be determined by $r(\theta)$ and $x(\theta)$. The gauge field on D5 can also be written as
 $A_t(\theta)$ for symmetry. The action of D5 brane is given by
 \bea\begin{split}
 S=T_5\Omega_4r_+^4\int \dd t \dd\theta &\sin^4\theta\\
 \biggr\{-&e^{\phi\over 2}\sqrt{\biggr[\biggr(1-{r_0^4\over r^4}\biggr)r^2+r'^2+
 \biggr(1-{r_0^4\over r^4}\biggr){r^4\over r_+^4}x'^2\biggr]-F_{\theta t}^2}
 +4A_t \biggr\},
 \end{split}
 \eea where $\Omega_4=8\pi^2/3$ is the volume of unit four sphere. To obtain the configuration of D5 brane, we should
 solve the gauge field at first. The equation of motion turns to be
 \bea
 \p_{\theta} D=-4\sin^4\theta.
 \eea The solution to the above equation is
 \bea
 \begin{split}
 D(\nu,\theta)={3\over 2}(\sin&\theta\cos\theta-\theta+\nu\pi)+\sin^3\theta\cos\theta,\\
 &0\leq\nu={k\over N_c}\leq1,
 \end{split}\eea where $k$ denotes the number of Born-Infeld strings emerging from south pole of S$^5$. More details
 about this solution can be found in~\cite{Ghoroku:2008na}. To eliminate the gauge field in favor of $D$, we shall transform
 the original Lagrangian to obtain an energy functional of the embedding function as follows
 \bea\label{rxL}
 \begin{split}
 \mathcal {H}
 =T_5\Omega_4r_+^4\int \dd\theta e^{\phi\over 2}\sqrt{\biggr[\biggr(1-{r_0^4\over r^4}\biggr)r^2+r'^2+
 \biggr(1-{r_0^4\over r^4}\biggr){r^4\over r_+^4}x'^2\biggr]}\times\sqrt{D^2+\sin^8\theta}\;.
\end{split}
 \eea
In order to find the configuration of D5 brane, one must extremize $\mathcal {H}$, with respect to $r(\theta)$ and
$x(\theta)$ respectively. A closed solution of D5 is argued as a physical baryon vertex. More discussion about these
solutions can be found in~\cite{GI0801}.
 \begin{figure}[t]
\centering
  \includegraphics*[width=0.4\columnwidth]{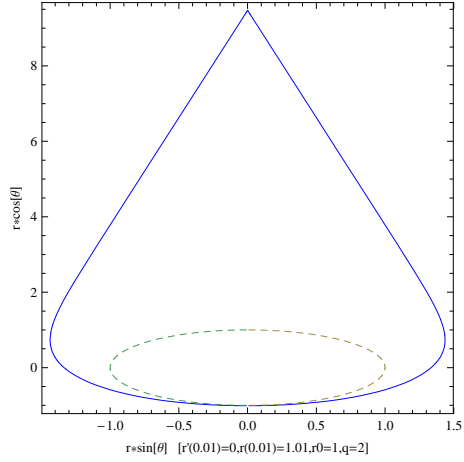}
  \caption{\small Baryon vertex configuration.}\label{bv}
\end{figure}
By solving equation of motion for $r(\theta)$ and $x(\theta)$ in (\ref{rxL}), we can find different kinds of solutions
for baryon vertex.

\subsection{Baryon vertex solutions} Note that point vertices in real spacetime
correspond to $x'(\theta)=0$. Thus the new action in (\ref{rxL}) turns to be \bea
 \begin{split}
 \mathcal {H}
 =T_5\Omega_4r_+^4\int \dd\theta e^{\phi\over 2}\sqrt{\biggr[\biggr(1-{r_0^4\over r^4}\biggr)r^2+r'^2\biggr]}\times\sqrt{D^2+\sin^8\theta}\;.
\end{split}
 \eea If $q=0$, the gravity
theory is the usual AdS black hole. In that case, vertex D5 brane with a DBI+CS action can not have a closed solution.
Here, we choose $q>0$(or some critical value) to keep closed D5 solutions, generally as in Fig \ref{bv}.  The solutions
are independent on $r_+$, if $x'(\theta)=0$. Only two parameters $q$ and $r_0$ determine behaviors of solutions. When
we choose suitable parameters $q=2,r_0\in(0.1,0.689)$, the vertex brane solutions can have four kinds of typical
behaviors if we choose different initial $r(0)$. Four kinds of typical behaviors correspond to four different kinds of
configurations of baryon. From these solutions, we see that there is always a singularity in $r_e=r(\pi)$, if we give
 initial conditions \bea\label{ic} r'(0)=0,\quad r(0)=C, \eea
where $C$ is a constant.

\subsection{Force balance condition} Adding fundamental
strings can help to eliminate this singularity and keep charge conserved. For simplicity, consider that $N_c$
fundamental strings all attach the north pole of S$^5$, which means $\nu=0$. $N_c$ static quarks are arranged on a
circle in ($x_1,x_2$) space, whose coordinates can also be written as ($\rho,\alpha$). By the following consistent
ansatz that describes the embedding fundamental strings \bea
 \tau=t,\quad \sigma=r,\quad \rho=\rho(r),
\eea we write the string action \bea
 S_F={1\over 2\pi\alpha'}\int \dd t \dd re^{\phi\over 2}\sqrt{\biggr(1+{r^4-r_0^4\over r_+^4}\rho'^2\biggr)}={1\over 2\pi\alpha'}\int \dd t
\dd r \LL_F\;. \eea To eliminate the singularity of cusp of D5 brane
at $r_e$, one needs force balance conditions. One force balance
condition in $\rho$ direction is satisfied for central symmetry.
Another force balance condition in $r$ direction is given by \bea
\label{FBC}
  N_c\biggr\{\LL_F -  \rho' {\p \LL_F \over \p \rho'}\biggr\}\biggr|_{r_e} = 2\pi \alpha'{\p \mathcal {H} \over \p
   r_e}\;. \eea The left hand of equation ( \ref{FBC} ) is the upward force of string and the right hand is the downward force of
   brane. The balance point is the singularity of vertex solution.
\section{Holographic melting}
\subsection{General analysis of screening length}
In the simplest example of the AdS/CFT correspondence provided by the duality between $\mathcal {N}=4$ supersymmetric
Yang-Mills theory and classical gravity in AdS$_5\times$S$^5$, external quarks can be introduced by fundamental strings
hanging from the boundary. At nonzero temperature, the potential between an external quark and an antiquark has
discontinuous behavior when their separation $L=L_s$. It's argued that the lower energy branch corresponding to larger
$r_e$ stands for real baryons, and $L_s$ is defined as ``screening length''. If quark separation $L>L_s$, the
quark-antiquark bound state will dissociate or melt in the medium. A similar screening length for baryons with multi
quarks waits to be analyzed. We shall calculate the screening effect for our baryon configuration. For the given
initial condition (\ref{ic}), one can obtain the cusp position $r_e$ ( the end of string in the bulk ) by solving the
equation of motion for $r(\theta)$. For different values of $r(0)$, the cusp attached by fundamental strings can have
different $r$ positions in the bulk geometry. In a recent work~\cite{Zhou:2008vf}, we analyzed the screening length for
D4 vertex brane plus $N_c$ fundamental strings, and found that the lower energy branch corresponded to smaller $r_e$
but not larger one. This is a very interesting finding. To see what screening effect will happen, let us turn to our
analysis for the present baryon configurations.

We shall consider quarks moving in medium and rotating in a plane, corresponding to boosted and high spin hadron state
respectively.\footnote{Actually, the general shape of multi quarks is a sphere in 3D space for largest symmetry, we
argue that results of a circle analyse is typical.} A static and spin zero state can be obtained by simplification from
complex states ( with more finite quantum numbers ) directly. The medium wind will effect the vertex brane and
fundamental strings at the same time, and the vertex brane can not feel the rotating effect, because it is a central
point in the rotation plane for symmetry. In the present case, we consider quarks moving in $x_3$ direction and
rotating in ($\rho,\alpha$) plane. For simplicity, we shall stand in the rest frame of the baryon configuration. The
metric (2.1) can be boosted such that it describes a gauge plasma moving with a wind velocity $v$ in the negative
$x_3$-direction. The boosted metric is given by~\cite{Liu:2008bs} \bea
 e^{-{1\over 2}\phi}ds_{10}^2=-A\dd t^2+2B\dd t\dd x_3+C\dd x_3^2+{r^2\over r_+^2}(\dd\rho^2+\rho^2\dd\alpha^2)+{r_+^2\over r^2}{1\over f(r)}\dd r^2+r_+^2\dd\Omega_5^2
\eea where \bea
 A={r^2\over r_+^2}\biggr(1-{r_1^4\over r^4}\biggr),\quad B={r_1^2r_2^2\over r^2r_+^2},\quad C={r^2\over r_+^2}\biggr(1+{r_2^4\over r^4}\biggr),
\eea with \bea
 r_1^4=r_0^4\cosh^2\eta, \quad r_2^4=r_0^4\sinh^2\eta,\quad v=-\tanh\eta,\quad f=1-{r_0^4\over r^4}.
\eea In the boosted metric, baryon configuration will depend on
$\eta$. Both vertex brane and fundamental string solutions will be
different from the original ones with $\eta=0$. First, we pay
attention to vertex brane solutions. For point brane vertex, one
notes that $x^i$ is independent on $\theta$. The D5 brane action in
boosted metric is given by \bea \mathcal
{H}_\eta=T_5\Omega_4r_+^4\int \dd\theta e^{\phi\over
2}\sqrt{\biggr(r^2+f^{-1}r'^2\biggr)\biggr(1-{r_0^4\cosh^2\eta\over
r^4}\biggr)}\times\sqrt{D^2+\sin^8\theta}\;. \eea To obtain rotating
fundamental string configuration, we give the following consistent
ansatz of embedding function \bea
 \tau=t,\quad\sigma=r,\quad\alpha=\omega t,\quad\rho=\rho(r).
\eea Facing the wind in $x_3$ direction, quarks arranged on the circle in $x_1-x_2$ plane will keep staying in
$x_1-x_2$ plane and stand on a circle, because they all have the same force.  Then the rotating string action in the
boosted metric can be written as \bea \label{sL}\begin{split}\tilde{S}_F={\mathcal {T}\over
2\pi\alpha'}\int_{r_e}^{r_\Lambda}\dd r\tilde{\LL}_F\;,\end{split}\eea where the Lagrangian \bea
\tilde{\LL}_F=e^{\phi\over 2}\sqrt{{r^2\over r_+^2}\biggr(1-{r_0^4\cosh^2\eta\over
r^4}-\rho^2\omega^2\biggr)\biggr({r_+^2\over r^2f}+{r^2\over r_+^2}\rho'^2\biggr)}\;. \eea The force balance condition
between D5 brane and $N_c$ fundamental strings turns to be \bea\label{fbc}
  N_c\biggr\{\tilde{\LL}_F -  \rho' {\p \tilde{\LL}_F \over \p \rho'}\biggr\}\biggr|_{r_e} = 2\pi \alpha'\Sigma(r_e,\eta)\;.
\eea To calculate ${\p \mathcal {H}' \over \p   r_e}$, we rewrite $\tilde{\mathcal {H}}=\mathcal {H}_\eta$
as\footnote{We assume that the vertex brane can not feel the rotation.} \bea\begin{split}
 \tilde{\mathcal {H}}
&=T_5\Omega_4r_+^4\int_{r_i=r(0)}^{r_e} \dd re^{\phi\over
2}\sqrt{\biggr(f^{-1}+r^2\theta'(r)^2\biggr)\biggr(1-{r_0^4\cosh^2\eta\over
r^4}\biggr)}\times\sqrt{D^2+\sin^8\theta(r)}\;. \end{split}\eea
Considering $r$ as ``time'', the ``hamiltonian'' shows the force of
vertex brane \bea\Sigma(r_e,\eta)={3\pi\over
2}T_5\Omega_4{r_+^4\over r^2}e^{\phi(r_e)\over
2}{\sqrt{r_e^4-r_0^4\cosh^2\eta}\over
\sqrt{r^2\theta'^2+f^{-1}}}*{1\over f}.
\eea The left hand in equation(\ref{fbc}) shows the force of strings \bea
 \tilde{\LL}_F -  \rho' \p_{\rho'}\tilde{\LL}_F={1\over \tilde{\LL}_F(r_e)}{e^\phi\over f}\bigg(1-{r_0^4\cosh^2\eta\over r_e^4}\bigg). \eea
 To solve the equation of motion of strings, we need two initial conditions. One is known by
$\rho(r_e)=0$ for symmetry, and the other must be calculated by the
force balance condition (\ref{fbc}). To get the baryon radius in the
boundary, we define \bea
 L_q=\int_{r_e}^{r_\Lambda}\rho'(r)\dd r.
\eea For $\eta>0,\omega=0$, the string Lagrangian (\ref{sL}) contains no $\rho$ and one can solve for $\rho'$ from the
equation of motion of $\rho$ and express the baryon radius $L_q$ in terms of $r_e$ and $\eta$ as follows \bea
 L_q(r_e,\eta)=\int_{r_e}^{r_\Lambda}\dd r {K(r_e,\eta)r_+^4\over \sqrt{r^4-r_0^4}\sqrt{e^\phi(r^4-r_0^4\cosh^2\eta)-K^2(r_e,\eta)r_+^4}}
\eea where $K(r_e,\eta)$ is constant by the equation of motion of $\rho(r)$, $\p_{\rho'}\tilde{\LL_F}=K$, determined by
the force balance condition \bea K(r_e,\eta)={1\over r_+^2}\sqrt{(r_e^4-r_0^4\cos^2\eta)e^{\phi(r_e)}}\biggr/\sqrt{1+
\theta'^{-2}r_e^{-2}f^{-1}}\;. \eea For $\eta>0,\omega>0$, the equation of motion of $\rho$ is difficult to solve
analytically. One must search for the numerical result. In both cases, screening length is always defined as the
maximum value of $L(r_e,\eta,\omega)$ ( as function of $r_e$ ) \bea
L_s(\eta,\omega)=\textbf{Max}\bigg\{L(r_e,\eta,\omega),r_e\bigg\}. \eea
\subsection{$q$ dependence of $L_s$ }
 \begin{figure}[t]
\centering
  \includegraphics*[width=0.5\columnwidth]{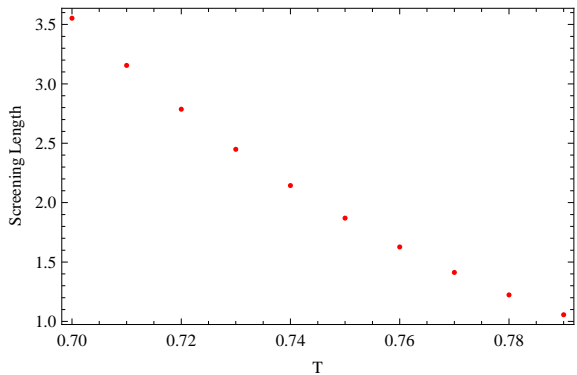}
  \caption{\small $T$ dependence of screening length. Usually, in plasma, $L_s$ is proportional to inverse Temperature.
  This Figure
shows that our result is consistent with this point. }\label{Sreening_T}
\end{figure}
 \begin{figure}[t]
\centering
  \includegraphics*[width=0.5\columnwidth]{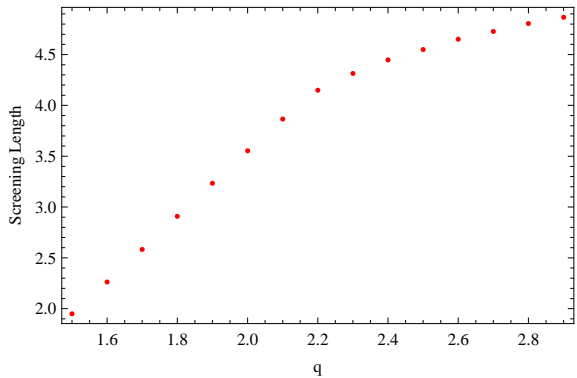}
  \caption{\small $q$ dependence of screening length.}\label{ls_q}
\end{figure}
In this gluon condensation background, we mainly modified the background with a dilaton factor $e^\phi=1+{q\over
r_0^4}\log{1\over 1-{r_0^4\over r^4}}$, and $q$ is the condensation parameter. If $q=0$, this background is same as the
AdS-black hole background. In that case, it is found that many properties of baryon screening are similar to those of
meson screening. The results are novel but the construction of baryon holographic model is not completed. Using the DBI
plus CS action of D5 vertex brane, one can not obtain a closed solution for baryon vertex in AdS-black hole background.
The usual physical explanation is that we can not get a closed baryon vertex in a deconfined gauge field background
which is dual to an AdS-black hole. In order to uncover the properties of thermal plasma by analyzing the heavy probe
behavior, we consider the gluon condensation modification in the present work. In this case, a closed vertex brane can
show us some interesting physical information. The solution of baryon vertex was firstly investigated in~\cite{GI0801},
and here we use the traditional method to add $N_c$ fundamental string by force balance condition. We go through the
traditional screening length calculation and find the screening length is modified by $q$ as shown in Figure
\ref{ls_q}. As $q$ increases, $L_s$ increases, which means that if we enhance the gluon condensation in the plasma, the
largest size of existing baryon increases. In another word, gluon condensate makes the plasma favor more baryons.
\subsection{$\eta$ dependence of $L_s$ }
 \begin{figure}[t]
\centering
  \includegraphics*[width=0.5\columnwidth]{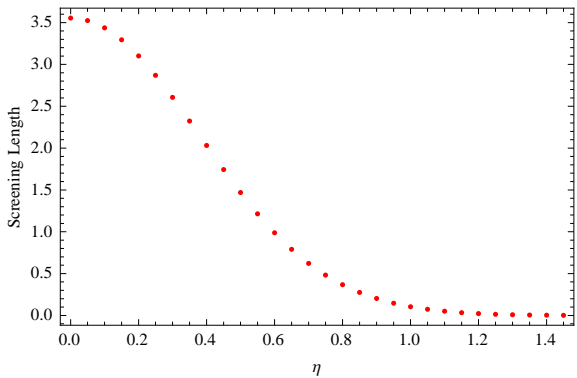}
  \caption{\small $\eta$ dependence of screening length. Here, we plot $L_s*\sqrt{\cosh\eta}$ vs $\eta$.}\label{ls_eta}
\end{figure}
Considering a probe of multi quark bound state with relative velocity $v$ in a plasma, the interaction between quarks
will be modified by $v$. The earliest work about this effect is~\cite{Liu:2008bs}. It's found that when $v$ goes to
one, $L_s$ is asymptotic to $L_0*(1-v^2)^{1/4}$. Here we use the same method to analyze the baryon probe in gluon
condensation background and find the result which is consistent to the result in~\cite{Liu:2008bs}. Note that the boost
parameter also effects the D5 vertex brane solution. In this work, when $\eta\geq1.6$($q=2,r_0=0.7$), since the
solution of $\eta$ dependent vertex brane action does not have a finite value at $\theta=\pi$ (it diverges), one can
not find a closed solution. The physical explanation is that the brane+strings configuration disappears above a
critical boost velocity, which implies that a heavy baryon with a very large velocity can not exist in the plasma or it
dissociates very quickly. This critical value of velocity depends on $q$ and $T$. This point is different
from~\cite{Liu:2008bs}, where the ideal model of baryon still holds at a infinite $\eta$ though the screening length is
almost zero. We can find the behavior when $0\leq\eta<1.6$. In this area, we observe that
$L_s*\sqrt{\cosh\eta}\sim$constant at a large $\eta$ in Fig \ref{ls_eta}. Our results are consistent with the
~\cite{Liu:2008bs}, and we observe that our screening length is larger than that in~\cite{Liu:2008bs} caused by the
gluon condensation $q$.
\subsection{$\omega$ dependence of $L_s$}
 \begin{figure}[t]
\centering
  \includegraphics*[width=0.5\columnwidth]{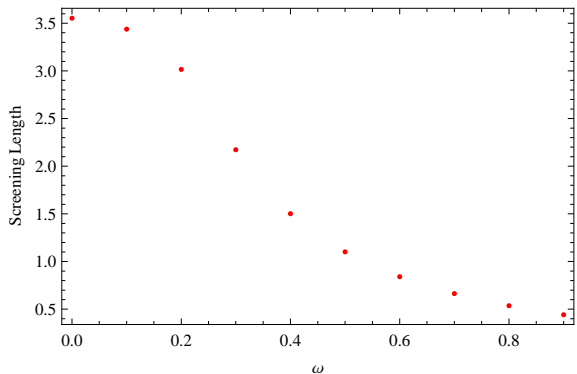}
  \caption{\small $\omega$ dependence of screening length.}\label{ls_omega}
\end{figure}

High spin meson can be presented with spinning string configuration~\cite{Peeters:2006}. The same kind of spinning
string configurations have been investigated in~\cite{Li:2008py} and the $\omega$ dependence behavior of screening
length has been found numerically. Here we do the same thing and find the result $L_s\sim \omega^{-1}$ in Figure
\ref{ls_omega} which is similar to the result in~\cite{Li:2008py}.
\section{Baryon mass and Melting analysis}
\begin{figure}[t]
\centering
  \includegraphics*[width=0.5\columnwidth]{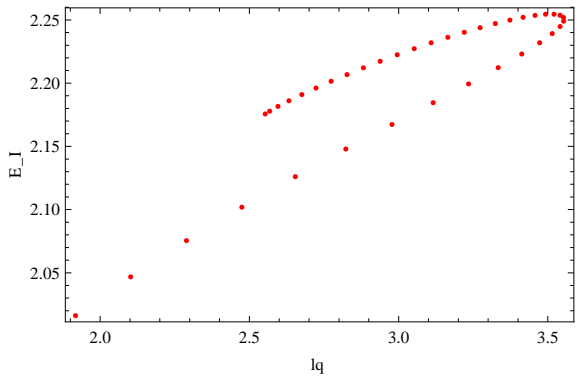}
  \caption{\small $l_q$ dependence of interaction potential.}\label{EI_lq}
\end{figure}

After calculating the screening length of holographic baryon probe, one wants to see how baryons dissociate in the
medium. As is well known, screening length is a property of the hot quark gluon plasma. If we want to judge whether a
baryon can survive in the plasma, we should compare the baryon radius with screening length of the plasma. Actually,
the life of real quark gluon plasma is very short, and the temperature of quark gluon plasma decreases very quickly.
Screening length $L_s$ always depends on temperature by $L_s\propto {\beta\over T}$, where $\beta$ is determined by the
properties of the real plasma. For a baryon with radius held fixed, as $T$ rises, it is going to melt. To make this
process clear, we should compute the interaction potential of baryon. The definition of baryon mass and interaction
potential are shown as follows. In a very general way, baryon mass is given by summation of the energy of $N_c$ strings
and the vertex brane as follows \bea\label{Etotal}
 E_{total}=N_cE_{string}+E_{D5},
\eea where the masses of string and vertex brane are given by\bea\label{stringenergy} E_{string}=\omega {\p \tilde{L}
\ov \p\omega}-\tilde{L}\;, \quad E_{D5}=\tilde{\mathcal {H}}\;,
 \eea
 Where $\tilde{L}=\frac {1}{2\pi \alpha'}\int_{r_e}^{r_{\Lambda}}\dd r \tilde{\LL}_F$ is the string Lagrangian.
 In order to obtain the interaction potential, one should analyze the free quarks
 case, in which $N_c$ strings hang from the boundary to $r_0$ and
 compact D5 brane almost wrapped on the $r=r_0$ contributes zero energy.
 Interaction potential is given by subtracting the energy of the free strings
. The radial distance
 of $r_0$ and boundary is $r_{\Lambda}-r_0$,  the mass of free quark is given respectively
 \bea
  E_q={1\ov 2\pi\alpha'}\int_{r_0}^{r_\Lambda}e^{\phi/2}\dd r\;
 \eea
 Then the interaction potential of baryon is obtained by
 \bea
  E_I=E_{total}-N_cE_q\;.
 \eea
Concretely, interaction potential can be written by \bea\begin{split}
 E_I={N_c\over 2\pi\alpha'}\int_{r_e}^{r_\Lambda}\dd r {e^\phi\over \tilde{\LL}_F}({1\over f}+{r^4\rho'^2\over r_+^4})(1-{r_0^4\cosh^2\eta\over
 r^4})+\tilde{\mathcal {H}}-{N_c\over 2\pi\alpha'}\int_{r_0}^{r_\Lambda}e^{\phi/2}\dd r
\end{split}\eea
Plotting the $E_I-L_q$ relation, as shown in Fig \ref{EI_lq}, we see that there exists two branches for each curve,
which shows that there are two energy states for fixed quark separation. We choose the low energy branch and argue that
it corresponds to the real physical baryon states. Similar curves were obtained for mesons and baryons in the usual
AdS$_5\times$S$^5$ background and other different backgrounds~\cite{Li:2008py,Zhou:2008vf}. One indeed can calculate
screening length and their dependence on speed and spin, but we can not give the exact prediction of the size and
energy of any heavy bound states. In our present gluon condensate case, we will find that the mass of baryons has
interesting behaviors and we hope it can give useful predictions for experiments~\cite{futurework}. We can also compare
baryons with mesons in gluon condensation background.
 \begin{figure}[t]
\centering
  \includegraphics*[width=0.5\columnwidth]{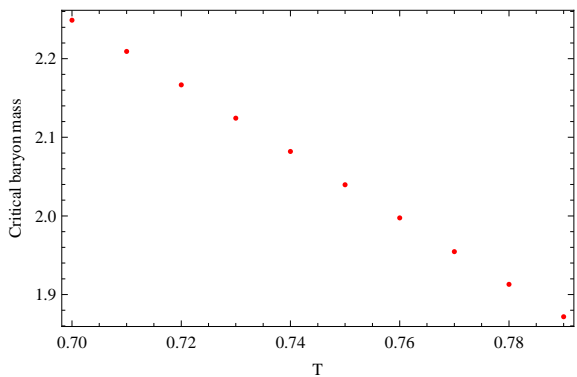}
  \caption{\small $T$ dependence of critical baryon mass. Since we have no bound state among the baryon configurations.
   We can just pick some special baryon mass and find how it depends on Temperature. Here we choose the critical mass corresponding to the screening point. The mean of this mass
is the highest energy of physical baryons surviving in the plasma. From this figure, we can see that mass is also
proportional to inverse Temperature.}\label{Mass_T}
\end{figure}

\section{Conclusion and discussion}
How to understand confinement and calculate hadron spectrum are considered as two biggest problems in QCD(or
non-perturbative QCD exactly). So far we know little about the non-perturbative world and have almost no general
powerful tool to study the strongly coupling problem. AdS/CFT correspondence, which is usually called gauge/gravity
duality in general, is believed as a useful framework to study these problems. In the experiment side, it is found that
there exists a QGP(quark gluon plasma) state in RHIC, which is a strongly coupled quark and gluon thermal state like a
fluid, investigated in many
works~\cite{Dusling:2008tg}~\cite{Erdmenger:2008rm}~\cite{Cai:2008in}~\cite{Cai:2008ph}~\cite{Figueras:2009iu}~\cite{Bhattacharyya:2008jc}~\cite{Rangamani:2008gi}.
How to describe this QGP and understand the strongly coupled behavior is still a problem, though it is very useful for
solving confinement and hadron spectrum problem. It is believed that heavy quark bound state can be alive in QGP,
including J/$\psi$ meson and some multi-quark bound states~\cite{de Forcrand:2000jx}. We call these multi-quark bound
states baryons, though they may be different from baryons we see when they survive within QGP. Using meson or baryon as
a probe is the simplest method to study the properties of the strongly coupled quark gluon state.

In the gauge/gravity duality framework, we calculate properties of the probe in the classical gravity background. From
the strong/weak duality, we know these results are always suitable for the probe in the strongly coupled background in
the field side. A lot of works have been done on the meson spectrum and meson melting process in different
gauge/gravity systems~\cite{Erdmenger:2007ja,Peeters:2006,Kobayashi:2006sb}.

In the present paper, we study holographic baryon probe with DBI + CS D5 vertex brane plus $N_c$ fundamental strings in
$N_c$ D3 branes background with gluon condensation at finite temperature. We investigate properties of this
configuration in a dilaton deformed AdS$_5\times$S$^5$ background \footnote{Besides the configuration considered in the
present work, where the D5 brane is pointlike along the gauge theory directions, one might envision the D5 growing Nc
(or fewer) spikes (i.e., Born-Infeld strings) which extend some distance along $\rho$ and then end at cusps, where they
are then continued by strings. Or perhaps the spikes could even extend all the way to the boundary, with the
fundamental strings shrinking down to zero length. Needless to say, it would be very difficult to find such solutions
explicitly. We ignore them purely in the construction of heavy baryon model in plasma for simplicity.}.
 We find that for most values of temperature $T$ and gluon condensation parameter $q$, there always
 exists a screening length $L_s$. The relation $L_s\sim{1\ov T}$ has been checked. We give the $q$ dependence of $L_s$.
 We also calculate the boost velocity $v=-\tanh\eta$ and angular velocity $\omega$ dependence of
$L_s$ for a baryon probe, which are consistent to those dependence relations in the point brane plus strings case, and
find that the usual relations have been modified by $q$. We also calculate the mass of baryon and find $T$ dependence
of baryon mass.

Another interesting way to investigate the baryon configuration in gauge/gravity duality is through the finite quark
density. The chemical potential of a finite quark density was introduced as a time component of the U(1) gauge filed on
flavor brane~\cite{Kobayashi:2006sb}. Finite quark density affect embedding of the flavor brane, as well as the phase
transition corresponding to meson dissociation. Following the present paper we can study the chemical potential
dependence of the baryon mass.

\section*{Acknowledgments}
Y.Zhou would like to acknowledge helpful discussions with Shi Pu and thank Pu for numerical help.

\end{document}